\documentclass[letterpaper, 10 pt, conference]{ieeeconf}
\IEEEoverridecommandlockouts                              

\usepackage{amsmath,amssymb}
\usepackage{mathtools}
\usepackage{booktabs}
\usepackage{array}
\usepackage{hyperref}
\usepackage{cite}
\usepackage{graphicx}
\usepackage{enumerate}
\usepackage{comment}
\usepackage{tabularx}
\renewcommand{\baselinestretch}{0.98}
\newcommand{\Real}{\mathbb{R}}
\newcommand{\norm}[1]{\left\|#1\right\|}

\begin{document}


\title{Stable but Unsafe: Agent-Driven Cyber-Physical Systems Under Gain Manipulation Attacks}

\author{Ali~Eslami and Jiangbo~Yu%
\thanks{Ali Eslami{\tt\small (ali.eslami@mcgill.ca)} and Jiangbo Yu {\tt\small (jiangbo.yu@mcgill.ca)} are with the department of Civil
Engineering, McGill University, Montreal, QC, Canada.}%
}

\maketitle
\begin{abstract}
AI agents are increasingly being connected to Cyber-Physical Systems (CPS) to generate or modify control-relevant parameters at runtime, including feedback gains, cost weights, and reference signals. These updates create a \emph{parameter channel}: a pathway between the agent and the controller that is structurally distinct from classical sensor and actuator channels. Among the parameters carried by this channel, feedback gains are especially high-leverage: under linear state feedback, a single gain matrix determines closed-loop eigenvalue placement for the entire system. Consequently, malicious gain updates can reshape the closed-loop dynamics without producing the signal-level inconsistencies targeted by residual-based monitors. We formalize this attack surface through a three-axis attacker model and a taxonomy of \emph{Gain Manipulation Attacks} (GMA). Two impact classes are identified: stability-margin erosion under sustained gain drift and transient amplification under one-shot gain replacement. We demonstrate that an attacker can drive the system past its safe physical operating limits while maintaining mathematical stability, proving that stability verification alone is insufficient to bound the physical impact. Using Bauer--Fike eigenvalue bounds and the Kreiss matrix theorem, we derive exact stealthiness conditions and worst-case impact certificates for each class. Finally, we propose preliminary detection directions and validate our framework through a vehicle lateral dynamics case study.
\end{abstract}

\vspace{-5pt}
\section{Introduction}
Recent agentic control architectures increasingly deploy AI agents within cyber-physical system (CPS) loops to generate parameters and signals such as feedback gains~\cite{costa2026maasc}, cost weights~\cite{maher2025llmpc}, and reference signals~\cite{wu2025instructmpc}. Rather than acting as passive or advisory overlays, these agents directly dictate the system's closed-loop behavior at runtime. In this paper, we define systems that feature such runtime, agent-driven modifications as \emph{agent-driven CPS}. This paradigm differs from classical adaptive, learned, or gain-scheduled control not because parameters vary, but because the updates are produced through an AI-agent execution stack whose outputs depend on stochastic model behavior, prompt structure, memory, tool use, or external data.

The interface carrying these runtime parameters is referred to here as the \emph{parameter channel}. It may transmit feedback gains, cost weights, reference signals, or other controller parameters and signals. Unlike sensor and actuator channels, which carry plant measurements and control inputs, the parameter channel modifies the controller mapping itself. As a result, a compromised parameter update can alter closed-loop dynamics without directly falsifying sensor measurements or actuator commands.

This architecture creates an attack surface that is not captured by conventional sensor- and actuator-channel attack models. The AI agent may be hosted in the cloud, on an edge processor, or on an onboard compute unit, but it typically runs on a general-purpose execution stack separate from the safety-certified controller it parameterizes. Two primary attack vectors are therefore especially relevant. First, an attacker may intercept and substitute agent outputs in transit between the agent and the controller. Second, the agent itself may be corrupted through memory poisoning, indirect prompt injection, or the manipulation of external context~\cite{greshake2023prompt}. In either case, a malicious update reaches the controller through an apparently legitimate configuration pathway.

Despite its direct influence on closed-loop behavior, the parameter channel remains largely unexplored as a distinct CPS attack surface. Existing CPS security research primarily studies attacks on plant signals, including false-data injection, replay, zero-dynamics, and covert attacks~\cite{ding2025topology,teixeira2012attack,eslami2025zero}. In those settings, the adversary manipulates measurements or control inputs, and detection and resilience mechanisms are typically designed around signal-level inconsistencies of the actuators and sensors. A related line of work studies data poisoning in data-driven control, where crafted perturbations to identification or training data can lead to degraded or destabilizing controller synthesis~\cite{russo2021poisoning,digge2026data,russo2023analysis}. These attacks primarily target the offline design or data pipeline. By contrast, the attacks considered here target controller parameters after deployment, during runtime operation.

Among the parameters carried by the parameter channel, feedback gains are particularly high-leverage targets. Under linear state feedback, a single gain matrix $K$ determines the closed-loop eigenstructure and directly shapes stability margins and transient response. We therefore focus on attacks that manipulate feedback gains at runtime, hereafter referred to as \textbf{Gain Manipulation Attacks (GMA)}. A key point is that such attacks need not destabilize the system to be harmful. A malicious gain may preserve closed-loop stability while eroding the stability margin or inducing large transient amplification. Thus, stability verification alone is insufficient to guarantee safety.

To the best of our knowledge, this paper provides the first systematic characterization of the parameter channel as a distinct runtime attack surface in agent-driven CPS. We formalize this attack surface through a taxonomy of GMA scenarios, define a stability-based notion of stealthiness, and derive closed-loop impact certificates for stability-margin erosion and transient amplification. Therefore, the contributions of this paper are as follows:
\begin{enumerate}
\item The parameter channel is formalized as a distinct runtime attack surface in agent-driven CPS. The three-axis attacker model of~\cite{teixeira2012attack} is extended to this channel, and a GMA taxonomy is developed by access point and temporal profile.

\item Closed-loop impact certificates are derived for the two main GMA impact classes: Bauer--Fike stability-margin certificates for sustained gain drift, and reachability, minimum-norm perturbation, and Kreiss-based transient-growth bounds for one-shot gain replacement. These results show that stability-preserving gains can still produce unsafe transient amplification.

\item Preliminary detection directions are proposed for man-in-the-middle and agent-side GMA by monitoring parameter-domain quantities, including stability margin and transient amplification, that are invisible to standard residual-based output monitors.
\end{enumerate}

\vspace{-5pt}
\section{System Model and Parameter Channel}
\label{sec:model}
Consider the following discrete-time LTI plant for the agent-driven CPS:
\begin{align}
x_{k+1} = A x_k + B u_k + w_k, \qquad y_k = C x_k + v_k,
\label{eq:plant}
\end{align}
with $x_k \in \Real^n$, $u_k \in \Real^m$, $y_k \in \Real^p$,
$w_k \sim \mathcal{N}(0,Q_w)$, $v_k \sim \mathcal{N}(0,R_v)$, where
$(A,B)$ and $(A,C)$ are controllable and observable.
The controller is governed by a feedback gain $K^{(i)} \in
\Real^{m \times n}$ issued by the agent at each update cycle $i$,
and we	
consider the state-feedback tracking case, $u_k = -K^{(i)}\hat{x}_k + r_k$,
where $\hat{x}_k$ is the observer state (defined subsequently) and $r_k$ is the reference
signal. The agent updates $K^{(i)}$ every $T_d \geq 1$ control steps;
between updates $K^{(i)}$ is held constant, inducing a switched LTI
structure. We assume that the updates satisfy a minimum dwell-time condition and therefore,
stability of the switched system is
guaranteed~\cite{liberzon2003switching}.
The parameter channel carries $K^{(i)}$ from agent to controller
at each agent update.
In a classical CPS, $K^{(i)}$ is fixed offline or updated by a verified,
deterministic supervisor with a bounded, auditable update rule; in an agent-driven CPS, $K^{(i)} = \phi(\xi^{(i)};\theta)$, where $\xi^{(i)}$ is the agent's runtime context and $\theta$ its parameters (weights, prompt). Note that the framework extends to other controllers that have gains and parameters as well, such as PID gains, MPC weights, etc.
Furthermore, a Luenberger observer reconstructs the state:
\begin{equation}
\hat{x}_{k+1} = A\hat{x}_k + Bu_k + L(y_k - C\hat{x}_k),
\label{eq:observer}
\end{equation}
with $L$ the observer gain.
In standard deployment, the observer sits at the controller node and
applies the same input $u_k$ as the plant; we refer to this as the
\emph{co-located observer architecture}.
The estimation error is then governed by:
$
x_{k+1}^e = (A - LC)x_k^e + w_k - Lv_k,
$
where $x_k^e \coloneqq x_k - \hat{x}_k$ is the estimation error.
Since the $Bu_k$ terms in~\eqref{eq:plant} and~\eqref{eq:observer} cancel, error dynamics are independent of $K^{(i)}$; residual-based monitors (e.g.,~$\chi^2$) are therefore blind to gain manipulation under the co-located architecture.
\vspace{-5pt}
\section{Threat Model and Impact Classes}
\label{sec:threat}
\vspace{-4pt}
\subsection{Attacker Model}
In this section, we extend the attacker model in \cite{teixeira2012attack} to parameter channel attacks.

\textbf{Definition 1}:
An attacker is characterized by three axes:
	I) \textbf{System knowledge}: what the attacker knows about the
	target, ranging from awareness that a parameter channel exists, to
	knowledge of the plant model, the nominal gain $K^{(i_0)}$
	and its operating range, the observer gain $L$, and the agent's
	internal structure $\phi(\cdot;\theta)$ including prompt or memory
	layout.
	II) \textbf{Disclosure resources}: the attacker's
	observation capability, ranging from no access, to observing the
	plant inputs and outputs, to full observation of the parameter
	channel, reading the information such as the communicated gain
	$K^{(i)}$.
	III) \textbf{Disruption resources}: a pair $(a, \mathcal{T})$ with
	access point $a \in \{\text{agent-side},\,\text{MITM}\}$ and target
	set $\mathcal{T} \subseteq \{K^{(i)}, r_k, \text{weights}, \ldots\}$
	denoting the attacked parameter components.
	Agent-side attacks corrupt $\xi^{(i)}$ or $\theta$ (memory
	poisoning, indirect prompt injection) and these corruptions then affect the generation of $K^{(i)}$. MITM attacks substitute the parameters on the channel such as $K^{(i)}$, similar to False Data Injection (FDI)
	attacks on input and output channels of the
	plant~\cite{teixeira2012attack}.

This paper takes $\mathcal{T} = \{K\}$ as a first step; attacks on 
other parameter-channel components are left for future work.
Furthermore, we instantiate Definition~1 as follows: the attacker
knows the plant and nominal control law $(A, B, K^{(i_0)})$ and can
read the parameter channel. Furthermore, depending on the access point, it can
write to the parameter channel (MITM) or corrupt the agent via memory
poisoning or indirect prompt injection (agent-side).
\vspace{-5pt}
\subsection{Attack Scenarios, Stealthiness, and Impact Classes}
\label{subsec:attack_scenarios}
We consider three concrete GMA scenarios covering
agent-side and MITM access. As established in Section~\ref{sec:model}
under the co-located observer architecture, residual-based monitors
cannot detect parameter-channel attacks. We therefore define stealthiness through a runtime stability check: both the
agent and the controller verify that any gain $\tilde{K}^{(i)}$ to be
issued or applied satisfies $\rho(A - B\tilde{K}^{(i)}) < 1$ before
it takes effect, where $\rho(\cdot)$ denotes the spectral radius of a
matrix, i.e., its largest eigenvalue in absolute value. An attack is
\emph{stability-stealthy} if the corrupted gain passes this check at
every agent update $i \geq i_0$:
\begin{equation}
\rho(A - B\tilde{K}^{(i)}) < 1, \quad \forall\, i \geq i_0.
\label{eq:stealthy}
\end{equation}
The agent-side check catches corrupted gains before they are sent; 
the controller-side check catches them before they are applied, 
forming a natural two-layer defense, the attack passing whichever check
applies at its access point.
The restriction is deliberate: an unstable gain is caught immediately, 
and the nontrivial question, whether an attacker confined to stabilizing gains can still cause dangerous behavior, is answered affirmatively below.
Note that as established in Section~\ref{sec:model}, the residual 
$y_k - C\hat{x}_k$ is independent of $K^{(i)}$ under the co-located 
observer architecture, and therefore carries no information about the 
gain, however large the resulting transient deviation becomes.

We consider the following attack scenarios for GMA:

\textbf{S1 -- One-shot gain jump (agent-side):}
	A single malicious prompt causes $\tilde{K}^{(i)} = K^{(i_0)} +
	\Delta K_1$ from attack onset $i_0$, where $\Delta K_1 \in
	\Real^{m \times n}$. This attack is stability-stealthy if
	$\rho(A - B(K^{(i_0)} + \Delta K_1)) < 1$.
	
\textbf{S2 -- Sustained gain drift (agent-side):}
	From attack onset $i_0$, at each agent update $i \geq i_0$, the
	issued gain drifts as $\tilde{K}^{(i)} = K^{(i_0)} -
	\alpha(i - i_0)E$, where $E \in \Real^{m\times n}$ is the drift
	direction matrix and $\alpha > 0$ is the drift rate.
	This attack is stability-stealthy if $\rho(A - B\tilde{K}^{(i)})
	< 1$ for all $i \geq i_0$, i.e., the drift remains within the
	stabilizing set at every update.
	
\textbf{S3 -- MITM gain replacement:}
	The attacker intercepts $K^{(i_0)}$ and substitutes it with
	$K^{(i_0)} + \Delta K_2$, $\Delta K_2 \in \Real^{m \times n}$, before it
	reaches the controller. This attack is stability-stealthy if
	$\rho(A - B(K^{(i_0)} + \Delta K_2)) < 1$. If the controller echoes the
	received gain to the agent, tampering is detectable by comparing sent and
	received values; otherwise the substitution is unobservable to the agent.

Stability-stealthy attacks are confined to the set of stabilizing gains, yet two distinct impacts remain possible within this set. \textbf{Definition~2:} \textbf{Class~I -- Transient amplification.} A stability-stealthy gain replacement $\tilde{K}=K^{(i_0)}+\Delta K_j$, $j\in\{1,2\}$, preserves closed-loop stability yet can cause a dangerous transient overshoot. With $\tilde{A}_{cl}\coloneqq A-B\tilde{K}$ denoting the perturbed closed-loop matrix, the worst-case transient gain $\Gamma(\tilde{A}_{cl})\coloneqq\sup_{k\geq0}\norm{\tilde{A}_{cl}^k}_2$ can substantially exceed $\Gamma(A_{cl}^{(i_0)})$, posing a safety risk even though the system eventually returns to steady state. Moreover, repeated one-shot replacements at successive update times can induce repeated transient overshoots while each applied gain remains stabilizing. \textbf{Class~II -- Stability-margin erosion.} Under sustained drift (S2), the spectral radius of the active closed-loop matrix $A_{cl}^{(i)}=A-B\tilde{K}^{(i)}$ drifts toward~$1$ while remaining below it at every step, progressively reducing the closed-loop stability margin.

    \vspace{-5pt}
\section{Analysis of Gain Manipulation Attacks}
\label{sec:attacks}

For each scenario, we characterize the worst-case stability-stealthy
attack subject to~\eqref{eq:stealthy}. For S1 and S3 (one-shot),
this means bounding the worst-case transient gain $\Gamma$ over the
reachable stabilizing set and identifying the structural conditions
under which it is maximized. For S2 (sustained drift), this means
finding the drift direction $E$ for a given rate $\alpha$ that
minimizes the Bauer--Fike certificate $i^{\rm BF}$, i.e., drives the
spectral radius of $A_{cl}^{(i)}$ toward~$1$ as rapidly as the
certificate guarantees. We use the convention that $i_0$ denotes the
attack onset, $K^{(i_0)}$ the nominal gain at onset,
$A_{cl}^{(i_0)} \coloneqq A - BK^{(i_0)}$ the nominal closed-loop
matrix at onset, and $\star$ denotes the BF-norm-optimal or
minimum-norm quantity as defined in each result
(e.g., $E^\star$ for the drift direction).
\vspace{-5pt}
\subsection{Gain Drift Certificate and Optimal Direction (S2)}
Under S2, substituting $\tilde{K}^{(i)} = K^{(i_0)} - \alpha(i-i_0)E$
into $A_{cl}^{(i)} = A - B\tilde{K}^{(i)}$ gives the active
closed-loop matrix at the $i$-th agent update:
\begin{align}
\begin{split}
A_{cl}^{(i)} &= A - B\bigl(K^{(i_0)} - \alpha(i-i_0)E\bigr)
\\&= A_{cl}^{(i_0)} + \alpha(i-i_0)\, BE.
\end{split}
\label{eq:Acli}
\end{align}
The key tool for analyzing how the eigenvalues of $A_{cl}^{(i)}$
evolve under this perturbation is the Bauer--Fike
theorem~\cite{golub2013matrix}: for a diagonalizable matrix, the
eigenvalues of a perturbed matrix cannot move farther than
$\kappa(V)\norm{\Delta}_2$ from the original eigenvalues, where
$\Delta$ is the perturbation and $\kappa(V)$ is the condition number
of the eigenvector matrix (defined subsequently in Assumption~1).
Applying this to \eqref{eq:Acli} yields a guaranteed lower bound
on the number of agent updates before instability onset, which we
call the \emph{certificate-of-survival} $i^{\rm BF}$: the system
is guaranteed stable for at least $i^{\rm BF}$ agent updates after
attack onset $i_0$.

\textbf{Assumption 1}: The nominal closed-loop matrix $A_{cl}^{(i_0)}$
is diagonalizable, i.e., $A_{cl}^{(i_0)} = V\Lambda V^{-1}$, where
$\Lambda$ is the diagonal matrix of eigenvalues, $V \in
\mathbb{C}^{n\times n}$ is the non-singular matrix of eigenvectors,
and $\kappa(V) \coloneqq \norm{V}_2\norm{V^{-1}}_2$ is the condition
number of $V$, measuring the degree of non-orthogonality of the
eigenvectors.

\textbf{Theorem 1}:
Under Assumption~1, for a given drift rate $\alpha > 0$ and drift
direction $E$ with $\norm{E}_F = 1$ ($||.||_F$ denotes the Frobenius norm), the active spectral radius
under S2 admits the Bauer--Fike bound
\begin{equation}
\rho(A_{cl}^{(i)}) \leq \rho(A_{cl}^{(i_0)})
+ \kappa(V)\cdot\alpha\cdot(i-i_0)\cdot\norm{BE}_2,
\quad\forall\,i\geq i_0,
\label{eq:bf_bound}
\end{equation}
Then, by denoting $i_* \coloneqq \min\{i \geq i_0 : \rho(A_{cl}^{(i)}) \geq 1\}$
as the first agent update at which the system becomes unstable,
the certificate-of-survival is obtained as
\begin{equation}
i_* \geq i_0 + i^{\rm BF},
\qquad
i^{\rm BF} \coloneqq \left\lfloor
\frac{1-\rho(A_{cl}^{(i_0)})}
{\kappa(V)\cdot\alpha\cdot\norm{BE}_2}
\right\rfloor
\label{eq:kcrit}
\end{equation}
A direction $E$ that maximizes $\norm{BE}_2$ over all
unit-Frobenius-norm directions (equivalently, minimizes $i^{\rm BF}$
and thus gives the fastest BF-certified instability onset) is
\begin{equation}
E^\star = v_{B,1} u_{B,1}^\top,
\label{eq:opt_E}
\end{equation}
where $u_{B,1}\in\Real^n$ and $v_{B,1}\in\Real^m$ are respectively
the leading left and right singular vectors of $B$ (from the SVD
$B = U\Sigma V_B^\top$, with $U\in\Real^{n\times n}$ and
$V_B\in\Real^{m\times m}$), giving
$\norm{BE^\star}_2 = \sigma_{\max}(B)$.
For $m=1$, $E^\star$ reduces to $B^\top/\norm{B}_2$.

\textbf{Proof}:
Set $\Delta_i\coloneqq\alpha(i-i_0)\,BE$.
By Bauer--Fike~\cite{golub2013matrix}, for any
$\lambda\in\sigma(A_{cl}^{(i)})$ there exists
$\lambda_j\in\sigma(A_{cl}^{(i_0)})$ with
$|\lambda-\lambda_j|\leq\kappa(V)\norm{\Delta_i}_2$.
Choosing $\lambda$ to realize $\rho(A_{cl}^{(i)})$ and applying the
triangle inequality:
\begin{align*}
\rho(A_{cl}^{(i)}) &= |\lambda| \leq |\lambda_j| +
\kappa(V)\norm{\Delta_i}_2 \\
&\leq \rho(A_{cl}^{(i_0)}) +
\kappa(V)\cdot\alpha(i-i_0)\cdot\norm{BE}_2.
\end{align*}
This gives \eqref{eq:bf_bound} directly; \eqref{eq:kcrit} follows
by solving for $i$.

With $B = U\Sigma V_B^\top$, set $E^\star = v_{B,1} u_{B,1}^\top$.
Then $\norm{E^\star}_F = \norm{v_{B,1}}_2\norm{u_{B,1}}_2 = 1$.
Computing: $BE^\star = \sigma_1 u_{B,1} u_{B,1}^\top$,
so $\norm{BE^\star}_2 = \sigma_1 = \sigma_{\max}(B)$.
For any $E$ with $\norm{E}_F = 1$, we have
$\norm{BE}_2 \leq \norm{B}_2\norm{E}_2 \leq \norm{B}_2\norm{E}_F
= \sigma_{\max}(B)$ (the second inequality uses $\norm{E}_2\leq\norm{E}_F$,
equality iff $E$ is rank-one). Since $E^\star$ attains this bound it
maximizes $\norm{BE}_2$ and hence minimizes $i^{\rm BF}$. The maximizer
is not unique: any $E=v_{B,1}w^\top$, $\norm{w}_2=1$, gives
$BE=\sigma_1 u_{B,1}w^\top$ and also attains $\sigma_{\max}(B)$. The choice $w = u_{B,1}$ in \eqref{eq:opt_E} is one example; 
any other unit vector $w$ gives the same $i^{\rm BF}$ but a different 
true onset $i_*$ (more clarifications in Remark~3). This completes the proof. \hfill $\blacksquare$

\textbf{Remark 1}:
Assumption~1 holds generically for systems with distinct eigenvalues,
which is the case for almost all physical systems. Near-repeated
eigenvalues lead to large $\kappa(V)$ and a conservative bound;
tighter certificates are available via pseudospectral
methods~\cite{trefethen2005spectra}.

\textbf{Remark 2}: For the \emph{defender}, $i^{\rm BF}$ is a 
guaranteed lower bound: the active mode is Schur-stable for all 
$i < i_0 + i^{\rm BF}$. For the \emph{attacker}, $i^{\rm BF}$ is 
not a prediction of instability onset; the actual $i_*$ may be 
much later (Remark~3).

\textbf{Remark 3}: A complementary direction is obtained via 
first-order eigenvalue perturbation: the direction maximizing 
the initial rate of growth of $|\lambda_d|$ is
$E^{**} = (B^\top\bar{y}_d)x_d^\top / 
(\|B^\top\bar{y}_d\|_2\|x_d\|_2)$,
where $y_d$, $x_d$ are the left and right eigenvectors of 
$A_{cl}^{(i_0)}$ associated with the dominant eigenvalue.
$E^{**}$ targets the dominant eigenvalue directly and has a 
steeper initial slope, whereas $E^\star$ maximizes the global 
perturbation norm $\|BE\|_2$. These objectives are not 
equivalent: over long drift horizons, $E^\star$ reaches 
instability sooner because the accumulated perturbation 
magnitude ultimately dominates a steeper initial slope, 
as confirmed in Section~\ref{sec:example}.
\vspace{-10pt}
\subsection{One-Shot Gain Replacement: Stability and Transient Bounds (S1, S3)}
\label{sec:oneshot}
For a one-shot gain replacement $\tilde{K} = K^{(i_0)} + \Delta K_j$,
$j \in \{1,2\}$, the closed-loop matrix becomes
$\tilde{A}_{cl} = A_{cl}^{(i_0)} - B\Delta K_j$.
Under the co-located architecture, the analysis is identical for
agent-side (S1) and MITM (S3) access; the two scenarios differ only
in attack and defense surface, not in dynamical consequences. MITM (S3) can be
defeated by cryptographic authentication of the parameter channel,
whereas agent-side (S1) survives authentication since the agent
itself produces the corrupted gain.
The following proposition characterizes three core properties
of any stability-stealthy one-shot attack: whether the perturbed
gain is certifiably stable, whether a target closed-loop matrix
is reachable through a gain change, and how large the resulting
transient can become.

\textbf{Proposition 1}:
	Let $\mathcal{R}_B \coloneqq \{BX : X \in \Real^{m\times n}\}$
	denote the set of $n\times n$ matrices whose columns lie in
	$\mathcal{R}(B)$, a linear subspace of $\Real^{n\times n}$ of
	dimension $\mathrm{rank}(B)\cdot n \leq mn$. Let
	$
		\mathcal{S}(\rho_{\max}, \tau) \coloneqq \bigl\{ M \in \Real^{n\times n} :
		\rho(M) \leq \rho_{\max},\;
		M - A_{cl}^{(i_0)} \in \mathcal{R}_B,\;
		\norm{B^\dagger(M - A_{cl}^{(i_0)})}_F \leq \tau \bigr\}
	$
	denote the reachable stability-stealthy set, where
	$\rho_{\max} \in [\rho(A_{cl}^{(i_0)}), 1)$ is a stealth budget,
	$\tau > 0$ is an attacker-chosen effort bound independent of
	\eqref{eq:stealthy}, and $B^\dagger$ denotes the Moore--Penrose
	pseudoinverse of $B$. Then:
	
	\begin{enumerate}
		
		\item \emph{(Stability)}: By Bauer--Fike applied to
		$\tilde{A}_{cl} = A_{cl}^{(i_0)} - B\Delta K_j$:
		\begin{equation}
			\kappa(V)\cdot\norm{B\Delta K_j}_2 < 1-\rho(A_{cl}^{(i_0)})
			\;\Longrightarrow\; \rho(\tilde{A}_{cl}) < 1.
			\label{eq:oneshot_bf}
		\end{equation}
		This condition is sufficient only; when it fails, stability
		must be verified directly.
		
		\item \emph{(Reachability)}: A target $\tilde{A}_{cl}^{\rm target}$
		is exactly reachable if and only if \vspace{-4pt}
		\begin{align}
			(I - \mathsf{P}_B)\bigl(\tilde{A}_{cl}^{\rm target} -
			A_{cl}^{(i_0)}\bigr) = 0,
			\label{eq:reach_condition}
		\end{align}
		where $\mathsf{P}_B \coloneqq BB^\dagger$ is the orthogonal
		projector onto $\mathcal{R}(B)$. In this case, the
		minimum-Frobenius-norm gain perturbation realizing it is
		\begin{equation}
			\Delta K_j^\star = B^{\dagger}
			\bigl(A_{cl}^{(i_0)} - \tilde{A}_{cl}^{\rm target}\bigr).
			\label{eq:deltaK_opt}
		\end{equation}
		When \eqref{eq:reach_condition} fails, \eqref{eq:deltaK_opt}
		returns the least-squares solution, realizing the nearest
		reachable matrix
		$A_{cl}^{(i_0)} + \mathsf{P}_B(\tilde{A}_{cl}^{\rm target} -
		A_{cl}^{(i_0)})$; the realized $\rho$ and $\Gamma$ must then
		be re-evaluated and \eqref{eq:stealthy} re-verified.
		When $B$ has full column rank, the reachable subspace has
		dimension exactly $mn$; for $m < n$, target matrices outside
		$\mathcal{R}_B$ cannot be realized.
		
		\item \emph{(Transient bound)}: For any $\tilde{A}_{cl} \in
		\mathcal{S}(\rho_{\max}, \tau)$, the worst-case transient gain
		satisfies the Kreiss bounds~\cite{trefethen2005spectra}
		\begin{equation}
			\mathcal{K}(\tilde{A}_{cl}) \leq \Gamma(\tilde{A}_{cl})
			\leq e \cdot n \cdot \mathcal{K}(\tilde{A}_{cl}),
			\label{eq:kreiss}
		\end{equation}
		where $\mathcal{K}(\tilde{A}_{cl}) \coloneqq
		\sup_{|z|>1}(|z|-1)\norm{(zI-\tilde{A}_{cl})^{-1}}_2$ is the
		Kreiss constant and $e \approx 2.718$ is Euler's number.
		When $\tilde{A}_{cl}$ is diagonalizable as
		$\tilde{V}\tilde{\Lambda}\tilde{V}^{-1}$, the tighter bound
		\begin{equation}
			\Gamma(\tilde{A}_{cl}) \leq \kappa(\tilde{V})
			\label{eq:kreiss_diag}
		\end{equation}
		applies, where $\kappa(\tilde{V}) \coloneqq
		\|\tilde{V}\|_2\|\tilde{V}^{-1}\|_2$.
		
	\end{enumerate}

\textbf{Proof}:
\emph{(Part 1.)} Set $\Delta = -B\Delta K_j$, so
$\norm{\Delta}_2 = \norm{B\Delta K_j}_2$. By Bauer--Fike
applied to $\tilde{A}_{cl} = A_{cl}^{(i_0)} + \Delta$,
$\rho(\tilde{A}_{cl}) \leq \rho(A_{cl}^{(i_0)}) +
\kappa(V)\norm{B\Delta K_j}_2 < 1$ whenever
\eqref{eq:oneshot_bf} holds.

\emph{(Part 2.)} The equation $B\Delta K_j =
A_{cl}^{(i_0)} - \tilde{A}_{cl}^{\rm target}$
has a solution iff each column of the right-hand side lies
in $\mathcal{R}(B)$; since
$(I-\mathsf{P}_B)(A_{cl}^{(i_0)} - \tilde{A}_{cl}^{\rm target}) = 0$
is equivalent to \eqref{eq:reach_condition}, the reachability
condition follows. The minimum-Frobenius-norm solution is
$\Delta K_j^\star = B^\dagger(A_{cl}^{(i_0)} -
\tilde{A}_{cl}^{\rm target})$ by the defining property of
the pseudoinverse~\cite{golub2013matrix}. Each column of
$B\Delta K_j$ lies in $\mathcal{R}(B)$, which has dimension
$\mathrm{rank}(B) \leq m$; with $n$ independent columns,
$\mathcal{R}_B$ has dimension $\mathrm{rank}(B)\cdot n \leq mn$,
with equality when $\mathrm{rank}(B) = m$.

\emph{(Part 3.)} The discrete-time Kreiss matrix
theorem~\cite{trefethen2005spectra} gives \eqref{eq:kreiss}.
For the diagonalizable case,
$\norm{\tilde{A}_{cl}^k}_2 =
\norm{\tilde{V}\tilde{\Lambda}^k\tilde{V}^{-1}}_2
\leq \kappa(\tilde{V})\norm{\tilde{\Lambda}^k}_2
= \kappa(\tilde{V})\rho(\tilde{A}_{cl})^k
\leq \kappa(\tilde{V})$
for all $k \geq 0$, since $\rho(\tilde{A}_{cl}) < 1$
implies $\rho(\tilde{A}_{cl})^k \leq 1$, giving
\eqref{eq:kreiss_diag}. \hfill $\blacksquare$

Proposition 1 characterizes a given attack's impact. Theorem 2 addresses the harder question: stealthiness imposes no bound on $\Gamma$, and an attacker engineering a non-normal closed-loop matrix can make $\Gamma$ arbitrarily large while passing the stability check.

\textbf{Theorem 2}:
	Let $\mathcal{S}(\rho_{\max}, \tau)$ be as in
	Proposition~1. For fixed $n$,
	$\rho_{\max} < 1$, and $\tau > 0$,
	$\sup_{M \in \mathcal{S}(\rho_{\max},\tau)}\Gamma(M)$
	is attained and finite. However, provided non-normal
	matrices are reachable in $\mathcal{S}$,
$
\sup_{M \in \mathcal{S}(\rho_{\max},\tau)} \Gamma(M)
\;\to\; +\infty
\quad \text{as } \rho_{\max} \to 1^-.
$
	The stability-stealthiness constraint alone does not
	bound Class~I impact when the stealth budget
	$\rho_{\max}$ is not fixed.

\textbf{Proof}: Since $M \in \mathcal{S}$ implies
$M - A_{cl}^{(i_0)} \in \mathcal{R}_B$, we have
$M - A_{cl}^{(i_0)} = BB^\dagger(M - A_{cl}^{(i_0)}) = BY$,
where $Y = B^\dagger(M - A_{cl}^{(i_0)})$ satisfies
$\norm{Y}_F \leq \tau$. Submultiplicativity gives
$\norm{M - A_{cl}^{(i_0)}}_F \leq \sigma_{\max}(B)\,\tau$,
bounding $\norm{M}_F$ and establishing boundedness of
$\mathcal{S}$. For closedness, each defining condition is
the pre-image of a closed set under a continuous map:
$\{M : M - A_{cl}^{(i_0)} \in \mathcal{R}_B\}$ is a closed
affine subspace; the effort condition is the pre-image of
$[0,\tau]$ under
$M \mapsto \norm{B^\dagger(M - A_{cl}^{(i_0)})}_F$;
and $\{M : \rho(M) \leq \rho_{\max}\}$ is closed since
eigenvalues vary continuously with matrix entries.
Hence $\mathcal{S}$ is compact, and since $\Gamma$ is
continuous on compact subsets of Schur-stable matrices,
$\sup_{M \in \mathcal{S}}\Gamma(M)$ is attained and finite.

For the unboundedness, a Jordan block $J_\rho$ at
eigenvalue $\rho \in (0,1)$ satisfies
$\norm{J_\rho^k}_2 \geq k\rho^{k-1}$
from the $(1,2)$ entry of $J_\rho^k$, so
$\Gamma(J_\rho) \geq \max_{k}\, k\rho^{k-1} \to \infty$
as $\rho \to 1^-$. Setting $\rho_{\max} = \rho$ and
$\tau \geq \norm{B^\dagger(J_\rho - A_{cl}^{(i_0)})}_F$
places $J_\rho \in \mathcal{S}(\rho_{\max},\tau)$
whenever $J_\rho - A_{cl}^{(i_0)} \in \mathcal{R}_B$,
which holds under the stated assumption. This completes the proof. \hfill
$\blacksquare$

\textbf{Remark 4}: Large $\Gamma$ is driven by
non-normality of $\tilde{A}_{cl}$, not by proximity
to the stability boundary: for any normal matrix $M$
(satisfying $MM^\top = M^\top M$),
$\mathcal{K}(M) = \Gamma(M) = 1$ regardless of $\rho(M)$,
so large transients require an ill-conditioned eigenvector
matrix. Since $\rho(\tilde{A}_{cl}) < 1$,
$\Gamma(\tilde{A}_{cl})$ is attained at a finite $k$,
making direct computation tractable; the bounds
\eqref{eq:kreiss}--\eqref{eq:kreiss_diag} serve as
analytical certificates when $\tilde{A}_{cl}$ is not
explicitly known, as in worst-case design problems.

\textbf{Remark 5:}
To construct a stability-stealthy gain replacement achieving a
prescribed transient amplification $\Gamma^*>1$, given
$\kappa^*\geq\Gamma^*$, choose eigenvalues $|\lambda_i|<1$ and
construct
$\tilde{V}=U\,\mathrm{diag}(\bar{\sigma},\ldots,\bar{\sigma}/\kappa^*)\,W^\top$
with orthogonal $U,W$, yielding $\kappa(\tilde{V})=\kappa^*$.
Set $\tilde{A}_{cl}^{\rm target}=\tilde{V}\,\mathrm{diag}(\lambda_i)\,\tilde{V}^{-1}$,
compute $\Delta K_j^\star$ via~\eqref{eq:deltaK_opt}, verify
\eqref{eq:reach_condition} and~\eqref{eq:stealthy}, and evaluate
$\Gamma$ directly. If $\Gamma<\Gamma^*$, increase $\kappa^*$
(monotonicity not guaranteed due to projection onto $\mathcal{R}_B$);
if still unmet, raise $\rho_{\max}$ toward~$1$ or move $\lambda_i$
closer to the unit circle.

\subsection{Preliminary Ideas on Detection of GMA}
\label{sec:detection}
GMA can happen through two approaches, each calling for a different detector. MITM attacks
(S3) are channel-integrity problem (issued and applied gains differ),
whereas agent-side corruption (S1, S2) is semantic: the corrupted
gain is faithfully issued, transmitted, and applied. The integrity tools of
classical CPS security (e.g., moving target \cite{ding2025topology} or auxiliary systems \cite{ eslami2025zero}) transfer to
the MITM face, exposing the mismatch between the issued gain and the observed
plant dynamics. However, a feedback channel from the controller to the agent, carrying the required information for detection of GMA is then necessary. However these approaches are blind to agent-side corruption, where issued and
applied gains coincide and no integrity check is violated. Detecting the latter
requires testing the effect of the gain, not its provenance. As the stealthiness condition~\eqref{eq:stealthy} is satisfied within the stabilizing set,
detection must operate in the parameter domain on $\{K^{(i)}\}$, monitoring two
quantities~\eqref{eq:stealthy} leaves free: the margin
$m^{(i)} \coloneqq 1-\rho(A_{cl}^{(i)})$ and the transient gain
$\Gamma(A_{cl}^{(i)})$ (Remark~4). A change-detection test on $m^{(i)}$ flags the
sustained erosion of Class~II (S2) and a test rejecting
$\Gamma(A_{cl}^{(i)}) > \Gamma_{\max}$ flags the non-normal amplification of
Class~I (S1/S3).
This approach matches the two impact classes of Definition~2,
so the monitor flags a stealthy GMA precisely when its impact is non-negligible.

\section{Numerical Example}
\label{sec:example}
As a representative safety-critical domain motivated by recent work on agentic vehicles and agentic transport systems~\cite{yu2025preparing,yu2025agentic}, we validate the theoretical results on the 4-state lateral--yaw--roll vehicle benchmark model of~\cite{gaspar2005design}. The state is $x_k = [v_y, \dot{\psi}, e_y, \phi]^\top$ and the input is $u_k = [\delta_f, \delta_r]^\top$, with the continuous-time model discretized by zero-order hold at $T_s = 0.05$~s. The agent update period is set to $T_d=20$ control steps, so one agent update corresponds to $T_dT_s=1$~s.
The nominal tracking LQR gain ($Q=\mathrm{diag}(1,1,10,1)$,
$R=0.01 I_2$), $K^{(i_0)}$, yields an initial closed-loop spectral radius
$\rho(A_{\text{cl}}^{(i_0)}) = 0.854$.

The resulting system diagnostics evaluate to $\kappa(V) = 9.62$,
$\sigma_{\max}(B) = 3.733$, and nominal transient envelope peak
$\Gamma(A_{\text{cl}}^{(i_0)}) = 3.07$. The BF-norm-optimal drift
direction $E^\star = v_{B,1}u_{B,1}^\top$ is formed from the leading
singular vectors of $B$, attaining $\norm{BE^\star}_2 =
\sigma_{\max}(B) = 3.733$.

\begin{figure}[t]
	\centering
	\includegraphics[width=0.95\columnwidth]{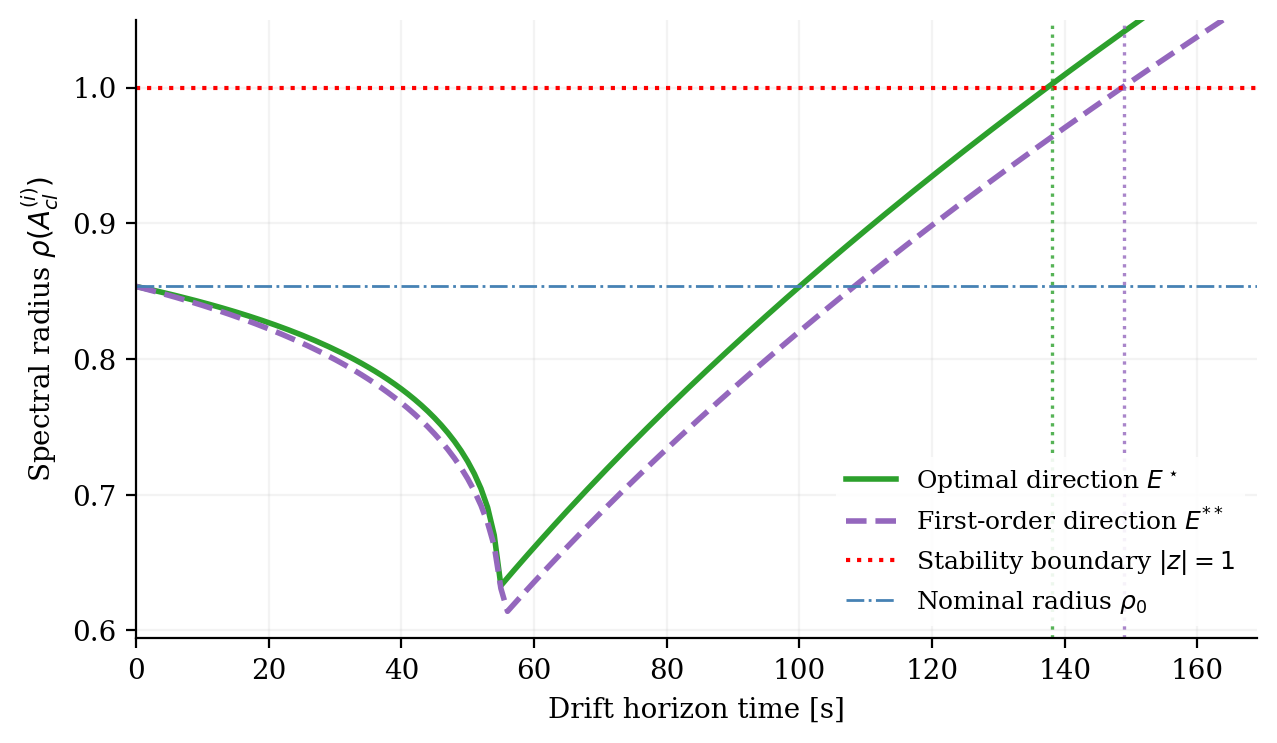}
	\caption{Class~II impact (Scenario S2, Theorem~1): closed-loop
		spectral radius $\rho(A_{cl}^{(i)})$ under $E^\star$ (solid) and
		$E^{**}$ (dashed). The BF-norm-optimal direction $E^\star$ reaches
		the stability boundary $|z|=1$ first at $i_*=138$ ($138$~s),
		versus $i_*=149$ ($149$~s) for the first-order direction
		$E^{**}$.}
	\label{fig:s2_drift}
	\vspace{-10pt}
\end{figure}

\emph{Scenario S2 (stability-margin erosion).}
S2 is instantiated with drift rate $\alpha = 2\times10^{-3}$ per update
along both candidate directions. Theorem~1 yields the conservative
Bauer--Fike certificate-of-survival $i^{\rm BF} = 2$ updates.
Applying first-order eigenvalue perturbation to $E^{**}$ predicts
instability at $i_*^{(1)}-i_0 = 116$ updates. Empirically,
instability occurs at $i_*(E^\star) = 138$ updates ($138$~s) and
$i_*(E^{**}) = 149$ updates ($149$~s), as shown
in Fig.~\ref{fig:s2_drift}.
Two effects predicted by the theory are visible. First, the
first-order estimate ($116$) is far less conservative than the
Bauer--Fike certificate ($2$); compared like-for-like against
the empirical onset of the \emph{same} direction $E^{**}$ ($149$),
it underestimates by a factor of $1.29$. Second, the
direction-optimality phenomenon of Remark~3 appears clearly:
although the first-order analysis ranks $E^{**}$ as the faster
direction, the BF-norm-optimal direction $E^\star$ breaches the
stability boundary sooner in practice ($138 < 149$), because over
a long drift horizon the growth of the global perturbation
magnitude $\norm{BE}_2$ dominates a steeper initial eigenvalue
slope. The transient dip near $55$~s in Fig.~\ref{fig:s2_drift}
corresponds to a non-dominant mode migrating inward before the
dominant pair turns and drives $\rho(A_{cl}^{(i)})$ monotonically
toward unity.

\begin{figure}[t]
	\centering
	\includegraphics[width=0.95\columnwidth]{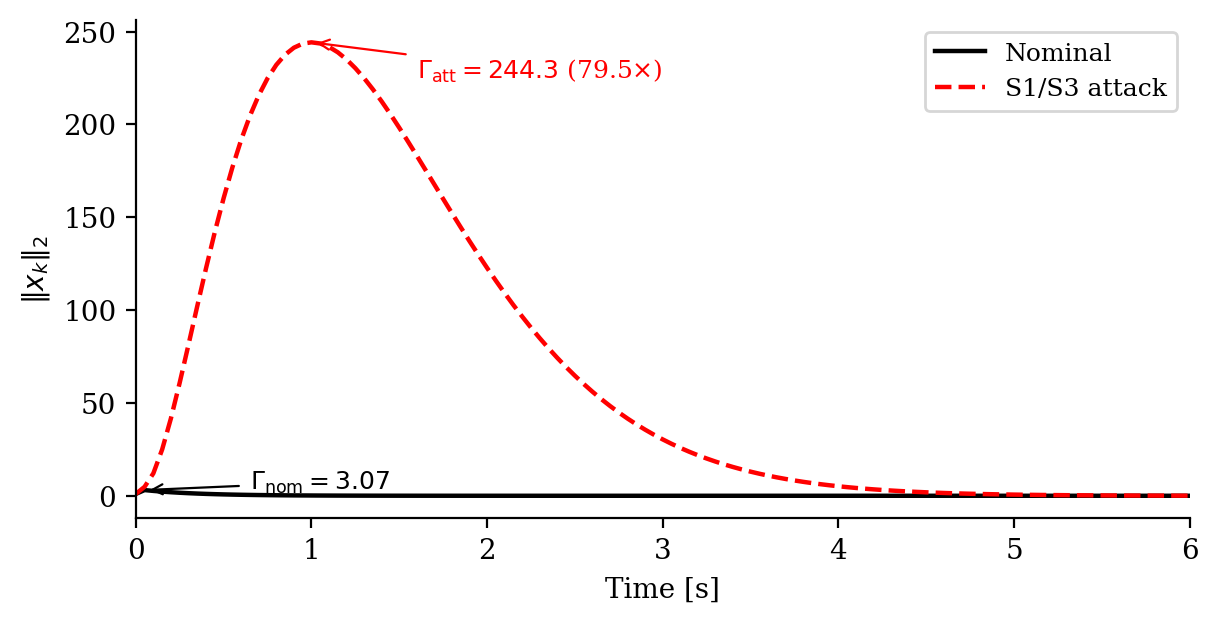}
	\caption{Class I transient amplification (S1/S3, Proposition~1 and
Theorem~2): state-norm trajectory $\lVert x_k\rVert_2$ under nominal
conditions and the one-shot gain attack. The system remains
asymptotically stable yet exhibits a transient peak
$\Gamma_{\text{att}} = 244.3$, a $79.5\times$ increase over the nominal
$\Gamma_{\text{nom}} = 3.07$, before decaying to the origin.}
	\label{fig:transient}
		\vspace{-10pt}
\end{figure}

\emph{Scenarios S1/S3 (transient amplification).}
A single malicious gain replacement $\tilde{K} = K^{(i_0)} + \Delta K_2$ is
obtained by a random search over $1000$ candidates in
$\mathcal{S}(\rho_{\max},\tau)$ with stealth budget $\rho_{\max}=0.91$
and effort bound $\tau = 1.0$, selecting the candidate with the largest
transient gain $\Gamma$; global optimality is not claimed and the
result simply represents a feasible high-impact attack:
$
\Delta K_2 = \begin{bmatrix}
0.4512 & -0.3943 & 0.1443 &  0.0330 \\
-0.5270 &  0.2428 & 0.4820 &  0.2238
\end{bmatrix},
$
with $\norm{\Delta K_2}_F = 1.00$ (comparable to $\norm{K^{i_0}}_F = 1.03$)
and $\norm{B\Delta K_2}_2 = 2.73$. The Bauer--Fike stability certificate
\eqref{eq:oneshot_bf} is uninformative here, since
$\kappa(V)\norm{B\Delta K_2}_2 = 26.3 \not< 1 -
\rho(A_{\text{cl}}^{(i_0)}) = 0.146$; direct spectral evaluation
nonetheless confirms closed-loop stability,
$\rho(\tilde{A}_{cl}) = 0.908 < 1$. Despite this margin, the transient
envelope grows from the nominal $\Gamma(A_{\text{cl}}^{(i_0)}) = 3.07$
to $\Gamma(\tilde{A}_{cl}) = 244.3$, a $79.5\times$ amplification whose
peak occurs at $k = 20$ control steps ($1$~s) after onset before the state
decays to the origin, as shown in Fig.~\ref{fig:transient}. Such
amplification can result in lateral deviations exceeding lane boundaries
or steering angles surpassing actuator saturation limits, triggering
anti-windup mechanisms and causing effective loss of lateral control
authority. The attacker
drives $\tilde{A}_{cl}$ into a strongly non-normal configuration, so a
stabilizing ($\rho < 1$), stealth-passing gain still produces a dangerously large transient. Because the resulting eigenbasis is
near-defective, the diagonalizable bound \eqref{eq:kreiss_diag} is far
too loose ($\kappa(\tilde{V}) \approx 1.4\times10^4$); the Kreiss bound
\eqref{eq:kreiss} is the meaningful characterization, with Kreiss
constant $\mathcal{K}(\tilde{A}_{cl}) = 125.4$ bracketing the transient
as $125.4 \leq \Gamma(\tilde{A}_{cl}) \leq e\,n\,\mathcal{K} = 1364$,
inside which the computed $\Gamma = 244.3$ indeed lies.
Table~\ref{tab:validation} summarizes all results.
The nominal $\Gamma$ reflects existing non-normality
under LQR; the attack exploits and amplifies this
while remaining stability-stealthy, confirming Theorem~2.

\begin{table}[t]
	\centering
	\caption{Analytical results versus simulation for stealthy 
		parameter channel attacks (S2: stability-margin erosion; 
		S1/S3: transient amplification).}
	\label{tab:validation}
	\renewcommand{\arraystretch}{1.1}
	\begin{tabularx}{\columnwidth}{@{}lXXX@{}}
		\toprule
		\textbf{Scenario} & \textbf{Metric} & \textbf{Theory} &
		\textbf{Simulation} \\
		\midrule
		S2 & survival certificate $i^{\rm BF}$  & 
		$\geq 2$ (certificate) & 
		$138$ ($E^\star$), $149$ ($E^{**}$); both $\geq 2$ \\
		S2 & $i_*-i_0$ for $E^{**}$  & 
		$116$ (first-order estimate) & 
		$149$ ($1.29\times$)$^{a}$ \\
		S1/S3 & $\rho(\tilde{A}_{cl})$ & $<1$ & $0.908$ \\
		S1/S3 & $\Gamma(\tilde{A}_{cl})$ &
		$[\,125.4,\,1364\,]$ (Kreiss) & 
		$244.3$ ($79.5\times$)$^{b}$ \\
		\midrule
		\multicolumn{4}{@{}l@{}}{%
			\footnotesize $^{a}$~Ratio: simulation/first-order 
			est.\ ($149/116$).}\\
		\multicolumn{4}{@{}l@{}}{%
		 \footnotesize $^{b}$~Amplification ratio: 
		$\Gamma(\tilde{A}_{cl})/\Gamma(A_{cl}^{(i_0)})$; 
		$\Gamma(A_{cl}^{(i_0)}) = 3.07$.}\\
		\bottomrule
	\end{tabularx}
	\vspace{-10pt}
\end{table}
\vspace{-5pt}
\section{Conclusion}
\vspace{-5pt}
\label{sec:conclusion}
We have characterized the parameter channel as a structurally distinct
attack surface in agent-driven CPS. The three-axis attacker model is
extended to the parameter channel, three canonical GMA scenarios and
two impact classes are established, and classical output-based
stealthiness is supplanted by a spectral radius condition. Formal
results include Bauer--Fike stability-margin certificates for gain
drift attacks, stealthiness and reachability conditions for one-shot
gain replacements, transient growth bounds via the Kreiss matrix
theorem, and a constructive attack design procedure. Furthermore, S1/S3
replacements can be repeated at successive agent updates, compounding
transient risk without triggering a stability alarm.

Future work includes solving the optimal gain drift attack as a
non-convex pseudospectral problem, extending the framework to
nonlinear plants via contraction theory, developing
parameter-domain detection mechanisms, designing a two-layer
parameter-integrity defense addressing both the channel and
the agent, analyzing multi-channel coordinated attacks that
extend the target set $\mathcal{T}$ beyond $\{K\}$, and
designing resilient controllers against this new attack surface.
\vspace{-15pt}
	\bibliographystyle{IEEEtran}
\bibliography{sample}  
\end{document}